\documentclass[10pt,fleqn,reqno]{article}
\usepackage[dvips]{graphicx}
\usepackage{epsfig}
\usepackage{psfig}
\usepackage{amssymb}
\begin{document}

\title{An associative network with spatially organized connectivity}
\author{Yasser Roudi$^{\S}$ and Alessandro Treves$^{\S,\P}$\\
\small $^{\S}$ SISSA, Cognitive Neuroscience sector, Trieste, Italy \\
\small $^{\P}$ NTNU, Centre for the Biology of Memory, Trondheim, Norway}

\maketitle
\begin{abstract}
We investigate the properties of an autoassociative network of
threshold-linear units whose synaptic connectivity is spatially
structured and asymmetric. Since the methods of equilibrium
statistical mechanics cannot be applied to such a network due to
the lack of a Hamiltonian, we approach the problem through a
signal-to-noise analysis, that we adapt to spatially organized
networks. The conditions are analyzed for the appearance of
stable, spatially non-uniform profiles of activity with large
overlaps with one of the stored patterns. It is also shown, with
simulations and analytic results, that the storage capacity does
not decrease much when the connectivity of the network becomes
short range. In addition, the method used here enables us to
calculate exactly the storage capacity of a randomly connected
network with arbitrary degree of dilution.
\end{abstract}

\section{Introduction}
Considerable theoretical and experimental evidence supports the
notion that cortical networks have been specialized in evolution
to serve a memory function. In particular the hippocampus, sitting
at the top of the cortical hierarchy \cite{Fel+91} is thought to
approximate a "pure" associative memory system - in the formation
of e.g. spatial memories in rats or episodic memories in humans
\cite{Squ92} - in the sense that the activity of individual units
is only meaningful in relation to previous activity, and not in
relation to the physical position of the units in the tissue
\cite{Tre+94}. At the core of the hippocampus, information from
different sources is associated together within the CA3 network,
and the pattern of activity corresponding to a memory item can be
conceived of as an arbitrary, randomly generated compressed
representation. In neocortical areas sitting lower in the
hierarchy, instead, memory operations still reflect the topography
informing synaptic connections, with the result that the activity
of a unit relates also to its position in the tissue. One can
identify of course many additional differences between memory
storage in the hippocampus and in the neocortex, e.g. differences
in time scales, but we focus here on two simple types of model,
that emphasize only the relationship between memory function and
the spatial organization ({\em i.e.}, the {\em geometry}) of the
connectivity. Both types of model implement an autoassociative
network with recurrent collateral connections, whose efficacy has
been modified during a training phase by associative plasticity (a
model Hebbian "learning rule").

In the first, `hippocampus' type of model it is assumed that
episodic memories or charts \cite{Sam+97} of the local environment
have been stored as patterns of neuronal activity distributed
throughout a network. Units in the network are labeled with an
index $i$, $i=1\dots N$, but the connectivity between the units, or
the probability that two units are connected, does not depend on
their indexes. The connectivity can in fact be complete, as in the
Hopfield model \cite{Hop82} and in its graded-response variants
\cite{Tre90}, or it can be sparse, but still independent of the
index, as in \cite{Som86} or in the highly diluted limit considered
by \cite{Der+87}. This type of model has been thoroughly analyzed in
terms of its storage capacity, yielding a relation between the
maximum number of pattern $p_c$ that can be turned into dynamical
attractors, i.e. that can be associatively retrieved, and the number
$C$ of connections per receiving unit. Typically the relationship
includes, as the only other crucial parameter, the sparseness of
firing $a$ \cite{Tre+91}, and it takes the form\begin{equation}
p_c\simeq {C\over a\log(1/a)}.\label{st_cap}\end{equation} Note that
the storage capacity calculation has been extended to the chart
model introduced by \cite{Sam+97}, leading to an estimate, parallel
to Eq.\ref{st_cap}, of the maximum number of charts that can be
stored given their sparsity $a$ and the number of connections $C$
\cite{Bat+98}. Although in a given chart units are arranged
topographically by their spatial selectivity, such an arrangement is
different from chart to chart and unrelated to any absolute "index"
- effectively there is a chart-specific index, randomly reshuffled
in each chart. Correspondingly, there is no absolute geometrical
structure to the connectivity, even though connection weights
reflect the storage of multiple charts.

Typically, the plasticity process, i.e. the modification of
connection weights that leads to the formation of attractors, is
not described in detail by mathematical models of the
autoassociative variety, but it is a widely held hypothesis that
in the hippocampal CA3 network attractors are formed by tuning the
synaptic efficacy of its recurrent collaterals with synaptic
plasticity mechanisms akin to LTP and LTD \cite{McN+87}. Very
similar mechanisms could operate in storing memory patterns in the
neocortex. Indeed, several reports on the observation of synaptic
plasticity in the isocortex contribute to this idea
\cite{Mar02,Buo98}. Thus memory storage could be mediated by the
same processes in the neocortex as in the hippocampus.

Yet, the first type of models reviewed above is inappropriate to
analyze memory retrieval in the cerebral cortex, because there one
has to take geometry into account. Both the local neocortical patch
and CA3, in terms of the {\em degree} of recurrent connectivity, can
be thought of as networks of extensively but sparsely connected
pyramidal cells \cite{Bra+91}, in the sense that each pyramidal
cells receives inputs from thousands of its neighbors, but those
represent only a fraction of the total neighbor population. While in
CA3, on the other hand, the probability of existence of a synapse
between two pyramidal cells does not change much with their physical
distance \cite{Ama+89}, in the neocortex on the contrary it does
depend on their distance. For instance one study \cite{Hel00} shows
that in layers II and III of mice visual cortex the probability of
connection falls off from $50\%$-$80\%$ for directly adjacent
neurons to $0\%$-$15\%$ at a distance of $500 \mu m$. A similar
distance dependence and spatially organized pattern of connectivity
could be observed in other isocortical areas, and this fact is what
is not considered in the type of models mentioned above, which makes
them inappropriate for neocortex.

Investigating a simple associative network model with a geometric
structure informing its connectivity is the purpose of this article.
We thus introduce and analyze an autoassociative network which is
comprised of threshold-linear units and includes a geometrical
organization of neuronal connectivity, meant as a simplistic model
of the type of organization of connections observed in the cortex.
The units in the model are therefore endowed with an index, that
refers to their physical position on an underlying substrate. For
simplicity, we consider periodic boundary conditions in either 1D (a
ring) or 2D (a torus). Connections are taken to be denser between
units close on the ring or the torus than between distant units.
Such connectivity structures have been considered extensively in
neural networks models of, for example, orientation selectivity
\cite{Ben+95} or head direction cells \cite{Tou98}, and have been
shown to lead to localized activity states ('bumps'), corresponding
to a specific orientation or head direction. These models do not
include an associative memory function. In addition to these models,
there have been studies on networks with non-geometric
connectivities but spatial correlations in the stored patterns
\cite{Mon93}. Here, we consider {\em both} geometry in the
connectivity and associative storage on the connection weights,
leading to network states than can be localized, {\em or} correlated
to randomly distributed activity patterns previously stored on the
weights, {\em or} both.

It is worth noting that it is not straightforward to apply to such
networks endowed with geometrical connectivity the methods from
statistical physics which were originally borrowed to solve models
like the Hopfield model \cite{Ami89}. These methods are based on the
existence of a Hamiltonian describing the steady, asynchronous
firing states of the system, which leads to a free-energy function
of a limited number of {\em order parameters}. One condition for the
existence of a Hamiltonian is that interactions between pair of
units be symmetric, i.e. the effect of a pre-synaptic unit on a
post-synaptic unit be exactly reciprocated. This obviously
presupposes symmetric connectivity (and identical weights in the two
directions); although it could also be taken to be a good first
approximation to networks with asymmetric connectivity\footnote{It
is worth-nothing that in a large variety of networks with graded
response units, the symmetric connectivity is just a necessary
condition for the existence of a Hamiltonian but not sufficient. It
ensures detailed balance and the existence of a Hamiltonian for
models with, for instance, binary neurons or monotonically
increasing analog response functions, but it does not suffice in a variety of
other models\cite{Kuh+91}}. Further, the standard procedure requires that
all variables that bear the index of individual units be averaged
out, to obtain a free-energy that depends solely on non-local,
extensive quantities, which can be assumed in turn to take narrowly
constrained values. As we shall discuss elsewhere, this {\em
self-averaging} property does not apply to networks with geometric
connectivity. To address these problems we develop an improved
version of the `self-consistent signal-to-noise analysis'
\cite{Shi+93}.

The paper is organized in the following way. In the second section,
we describe a model of an associative network of threshold-linear
model neurons with an arbitrary geometrical (and sparse)
connectivity. We then derive the self-consistent equations for the
order parameters that we define. We refer to these equations as the
mean-field equations. In the third section, we use these equations
to calculate the storage capacity of a network without geometry. We
recover the results previously found by using the replica method for
an `extremely diluted' network \cite{Tre91a} and also calculate the
leading deviations from this limit when the connectivity is less
sparse. This exercise yields insight useful later in analyzing the
geometrical model. In the fourth section, we study a one-dimensional
model in which we consider a probability distribution for the
connectivity. We study the behavior of the storage capacity and the
shape of the profiles of activity for such a network, via analytical
arguments and computer simulations. Conclusions are summarized at
the end, while details of the calculation are provided in 3
Appendices.

\section{Methods}
\subsection{Threshold-linear model}
Consider a network of $N$ units, in which the level of activity of
unit $i$ is represented by a variable $v_i\ge 0$. This variable can
be taken to represent the firing rate of the neuron averaged over a
short time window. We assume that each unit receives $C$ inputs from the
other units in the network. The thermodynamic limit $N\rightarrow
\infty$ and $C \rightarrow \infty$ is assumed. The specific
covariance 'Hebbian' learning rule we consider prescribes that the
synaptic weight between units $i$ and $j$ be given as:
\begin{equation}
J_{ij}=\frac{1}{Ca^2}\sum_{\mu=1}^p c_{ij}\left(\eta_i^{\mu}-a\right)\left(\eta_j^{\mu}-a\right),
\end{equation}
where $\eta_i^{\mu}$ represents the activity of unit $i$ in memory
pattern ${\mu}$ and $c_{ij}$ is a binary variable equal to $1$ if
there is a connection running from neuron $j$ to neuron $i$, and
$0$ otherwise. Each $\eta_i^{\mu}$ is taken to be a `quenched
variable', i.e. a given parameter, drawn independently from a
distribution $p(\eta)$, with the constraints $\eta \ge 0$,
$\langle \eta \rangle = \langle \eta^2 \rangle= a$, where
$\langle\rangle$ stands for the average over the distribution
$p(\eta)$. Here we concentrate on the binary coding scheme
$p(\eta)=a\delta(\eta-1)+(1-a)\delta(\eta)$, but the calculation
can be carried out for any probability distribution. As in one of
the first extensions of the Hopfield model \cite{Tso+88}, we thus
allow for the mean activity $a$ of the patterns to differ from the
value $a=1/2$ of the original model \cite{Tre90}. We further
assume that the input (local field) to unit $i$ takes the form:
\begin{equation}
h_i=\sum_{j\neq i}J_{ij} v_i+ b\left(\frac{1}{N}\sum_jv_j\right),
\end{equation}
where the first term enables the memories encoded in the weights
to determine the dynamics; the second term is unrelated to the
memory patterns, but is designed to regulate the activity of the
network, so that at any moment in time $x\equiv\frac{1}{N}\sum_i
v_i$ and $y \equiv \frac{1}{N}\sum_i v_i^2$ both approach the
prescribed value $a$ (which then parametrizes the {\em sparsity}
of the network activity \cite{Tre+91}). The activity of each unit
is determined by its input through a threshold-linear function:
\begin{equation}
v_i=F[h_i]=g(h_i-T_{thr})\Theta(h_i-T_{thr})
\end{equation}
where $T_{thr}$ is a threshold below which the input elicits no
output, $g$ is a gain parameter, and $\Theta(...)$ the Heaviside
step function. The exact details of the updating rule are not
specified further, here, because they do not affect the steady
states of the dynamics, and we take ``fast noise'' levels to be
vanishingly small, $T\to 0$. Discussions about the biological
plausibility of this model for networks of pyramidal cells can be
found in \cite{Tre+91,Ami91}, and will not be repeated here.

In order to analyze this network, we first define a set of order
parameters $\{m^\mu_i\}$, with $\mu=1\ldots p;  i=1\ldots N$, which
we call the {\it local overlaps}, as follows:
\begin{equation}
m^{\mu}_i=\frac{1}{C}\sum_j c_{ij}(\eta^{\mu}_j/a-1) v_j,\label{eq4}
\end{equation}
This is a natural choice for quantities that measure retrieval
while also taking into account the spatial structure of the
network, and hence the position dependence of the activity.

If we rewrite the local field $h_i$ defined above in terms of
these order parameters we have:
\begin{equation}
h_i=\sum_{\mu} \left(\eta_i^{\mu}/a-1\right) m_i^{\mu}- c_{ii}\alpha
(1/a-1) v_i\label{loc_field}+b\left(x\right)
\end{equation}
in which $\alpha=p/C$ is the storage load, and we have carried out
the average $\sum_{\mu}(\eta_i^{\mu}/a-1)^2\simeq p(1/a-1)$. We
will use this expression for the local field in the next section.

\subsection{Retrieval states and the mean-field equations}
A pattern $\mu$ is said to be retrieved if $\sum_i m^{\mu}_i=O(N)$.
Without loss of generality, we suppose that the first pattern is the
retrieved one and therefore $ m^{\nu}_i \ll m^{1}_i$ for $\nu \neq
1$ and any $i$. When one pattern is retrieved, the local field to
each unit can be decomposed into two terms. One is the {\em signal},
which is in the direction of keeping the network in a state with
large overlap with the retrieved pattern. The second term, which can
be called {\em noise}, contributes random interference. In
Eq.\ref{loc_field} the signal is nothing but the $\mu=1$ term in the
sum on the r.h.s., whereas the noise is the rest. The idea is to
calculate these terms as a function of the local overlaps with the
retrieved pattern. In other words we wish to express the r.h.s of
Eq.\ref{loc_field} solely as a function of $m_i=m^{1}_i$ and
$\eta^1_i$. If we do so, we can then express the activity of each
unit as a function of $m_i$, and by inserting it in the definition
of local overlaps, we will be able to find a self consistent
equation for the local overlap with the first pattern.

To proceed further, we define two more local order parameters
$\rho_i$ and $\gamma_i$ through the equation:
\begin{equation}
\sum_{\nu \neq 1}(\eta_i^{\nu}/a-1) m_i^{\nu}=\rho_i z+\gamma_i v_i,
\label{noise}
\end{equation}
where we take $z$ to have quenched-averaged standard deviation
unity. We then single out a generic pattern $\mu$ from the sum
over non-retrieved patterns, writing:
\begin{equation}
\sum_{\nu \neq 1,\mu}(\eta_i^{\nu}/a-1) m_i^{\nu}=\rho^{\mu}_i
z+\gamma^{\mu}_i v_i,
\end{equation}
and noting that, to leading order in $1/p$, $\rho^{\mu}_i \simeq
\rho_i$ and $\gamma^{\mu}_i\simeq \gamma_i$ are expected to be
independent of $\mu$. With this, we can write the activity of the
network as:
\begin{equation}
v_i=F[(\eta_{i}^{1}/a-1) m_{i}^{1}+(\eta_{i}^{\mu}/a-1)
m_{i}^{\mu}+\rho^{\mu}_i z+\gamma^{\mu}_i
v_i-c_{ii}\alpha(1/a-1)+b(x)-T_{thr}]
\end{equation}
from which $v_i$ can be found self consistently, as in
\cite{Bat+98}:
\begin{equation}
v_i\simeq G[(\eta_{i}^{1}/a-1) m_{i}^{1}+(\eta_{i}^{\mu}/a-1)
m_{i}^{\mu}+\rho_i^{\mu}z+b(x)-T_{thr}]
\end{equation}
Assuming that $\Gamma_i =\gamma_i-c_{ii}\alpha (1/a-1) < 1/g$
\footnote{We shall see later that this assumption is valid,
at least when one deals with diluted networks or very low storage
loads.} the function $G[x]$ takes the following form
for a threshold-linear unit:
\begin{equation}
G[x]=\frac{g}{1-g\Gamma}x \Theta(x).
\end{equation}
In the case of a non-geometric network, as discussed by Shiino and
Yamana \cite{Shi04} this factor $\Gamma$ equals minus the Onsager
reaction term, when one treats the network in the TAP equation
framework.

Now we expand the r.h.s. of the above equation for $v_i$ up to the
linear term in $m^{\mu}_i$ and insert the result in Eq.\ref{eq4}, to
get:
\begin{equation}
m_i^{\mu}=L_i^{\mu}+\sum_{j}K_{ij}^{\mu} m_{j}^{\mu} \label{mimu}
\end{equation}
where:
\begin{eqnarray*}
L_i^{\mu}&=&\frac{1}{C}\sum_{j}c_{ij}(\eta_j^{\mu}/a-1)G[(\eta_{j}^{1}/a-1) m_{j}^{1}+\rho_j^{\mu}z+b(x)-T_{thr}]\\
K_{ij}^{\mu}&=&\frac{c_{ij}}{C}(\eta_j^{\mu}/a-1)^2
G'[(\eta_{j}^{1}/a-1) m_{j}^{1}+\rho_j^{\mu}z+b(x)-T_{thr}].
\end{eqnarray*}
Solving the above equation for $m_i^{\mu}, \mu \neq 1$ and using
it in the expression defining $m_i\equiv m_i^{1}$ we get the
following self consistent (mean-field) equations (see Appendix I):
\begin{eqnarray}
\psi_{ij}&=&\sum_l K_{il}^{\mu}c_{lj}+\sum_{lt}
K_{il}^{\mu}K_{lt}^{\mu}c_{tj}+\ldots\nonumber\\
\Gamma_i&=&\alpha T_0 \psi_{ii}\nonumber \\
\rho^2_i&=&\frac{\alpha g^2 T_0^2}{C} \sum_{j}\left( c_{ij}+2 c_{ij}\psi_{ij}+\psi_{ij}^2\right)\times\nonumber\\
&&\langle\int^{+}Dz
\left((\frac{\eta_j}{a}-1)m_j+b(x)-T_{thr}-\rho_j z\right)^2
(1-g\Gamma_{j})^{-2}\rangle \label{mean_field}\\
m_i&=&\frac{g}{C} \sum_{j} c_{ij} (\eta_{j}/a-1)\times\nonumber\\
&&\int^{+} Dz \left((\frac{\eta_j}{a}-1)m_j+b(x)-T_{thr}-\rho_j z\right)(1-g\Gamma_j)^{-1}\nonumber\\
x&=&\frac{g}{N}\sum_j\langle\int^{+} Dz
\left((\frac{\eta_j}{a}-1)m_j+b(x)-T_{thr}-\rho_j
z\right)(1-g\Gamma_j)^{-1}\rangle. \nonumber
\end{eqnarray}
where $Dz=dz \frac{e^{-z^2/2}}{\surd 2 \pi}$ and the superscript
$+$ indicates that the integration has to be carried out in the
range where $(\frac{\eta_i}{a}-1)m_i+b(x)-T_{thr} > \rho_iz$. The new
order parameters $\psi$, $\rho$ have been defined in the
derivation of these equations in Appendix I.

\section{Diluted network without structure}
To proceed further let us first consider the case in which
there is no geometry and the $c_{ij}$'s are randomly generated
with probability $Pr\{c_{ij}=1\}=C/N$. In this case, in the definition of $\psi$
(see Eq.\ref{loop} in Appendix I) for the first sum on the r.h.s
we have:
\begin{equation}
\sum_l K_{il}^{\mu}c_{lj}=\frac{1}{C}\sum_{l} c_{il}c_{lj}
(\eta^{\mu}_l/a-1)^2 G'[(\eta_{l}^{1}/a-1)
m_{l}^{1}+\rho_l^{\mu}z+b(x)-T_{thr}].\nonumber
\end{equation}
If we replace the sum with an average over the distribution of
$\{c_{ij}\}$ and $\{\eta_i\}$ and neglect in this averaging any
correlation between the position of the unit and its activity
(this assumption will have to be reviewed, of course, in the
geometric case)
\begin{equation}
\sum_l K_{il}^{\mu}c_{lj}=\frac{C}{N} <(\eta/a-1)^2>_{\eta}
<G'[j]>_{\eta,z}=\frac{CT_0}{N}<G'[j]>_{\eta,z}\nonumber
\end{equation}
and, in fact, one notes that it can be written, to any order in
$n$:
\begin{equation}
\sum_{l,t}
K_{il}^{\mu}\left(K^{\mu}\right)_{lt}^{n-1}c_{tj}=
\frac{C}{N}\left[T_0<G'[j]>\right]^{n}_{\eta,z}=
\frac{C}{N}\Omega^n \nonumber
\end{equation}
where we have defined the quantity:
\begin{equation}
\Omega=T_0<G'[j]>_{\eta,z}=\frac{gT_0}{1-g\Gamma}<\int^+Dz>
\label{Omega_def}
\end{equation}
that we assume to self-average, {\em i.e.} not to depend on the
index $j$.

Using the above expression yields the steady state equations:
\begin{eqnarray}
\psi&=&\frac{C}{N}\left(\Omega+\Omega^2+\Omega^3+\ldots\right)\nonumber\\
\Gamma&=&\alpha T_0 \psi\nonumber\\
\rho^2&=&\alpha\left(\frac{gT_0}{(1-g\Gamma)}\right)^2 \left(1+2\psi+\frac{N}{C} \psi^{2}\right) \times\label{ss_eqs}\\
&&\langle\int^{+}Dz \left((\frac{\eta}{a}-1)m+b(x)-T_{thr}-\rho z\right)^2\rangle\nonumber\\
m&=&\frac{g}{1-g\Gamma} \langle\int^{+}Dz (\eta/a-1) \left((\frac{\eta}{a}-1)m+b(x)-T_{thr}-\rho z\right) \rangle\nonumber\\
x&=&\frac{g}{1-g\Gamma} \langle\int^{+}Dz \left((\frac{\eta}{a}-1)m+b(x)-T_{thr}-\rho z\right)\rangle.\nonumber
\end{eqnarray}

The fact that $\psi$ vanishes in the limit of $C/N
\rightarrow 0$ can be understood intuitively. The order parameter
$\psi$ is nothing but the contribution of the activity
reverberating in the loops of the network. When one considers a
highly diluted network, the number of such loops becomes
negligible, and they do not contribute to network dynamics. Thus
$\psi$ and $\Gamma$ vanish in this limit. This also makes
the above inequality $\Gamma_i < 1/g$ a valid assumption.
$\Gamma$ essentially measures the effect of the activity of each
unit on itself, after it has reverberated through the network, and
this effect becomes negligible when one deals with an extremely
diluted network.

We can then define the new variables $r=m/\rho$ and
$w=[b\left(x\right)-m-T_{thr}]/\rho$ and the following integrals,
which are functions of $r$ and $w$, as in \cite{Tre90}:
\begin{eqnarray}
A_2&=&\frac{1}{rT_0}\langle (\frac{\eta}{a}-1)\int^+Dz (w+\frac{r\eta}{a}-z)\rangle \nonumber\\
A_1&=&A_2-\langle\int^{+}Dz\rangle  \\
A_3&=&\langle \int^+Dz(w+\frac{v\eta}{a}-z)^2\rangle. \nonumber
\end{eqnarray}
By using this notation, as shown in Appendix II, one finds that:
\begin{equation}
\Omega=1-(A_1/A_2)
\end{equation}
and the remaining steady state equations can be reduced to:
\begin{eqnarray}
E_1(r,w)&=&A_2^2-\left(1+\frac{C}{N}\left(\frac{(2-\Omega) \Omega}{(1-\Omega)^{2}}\right)\right)\alpha A_3=0 \label{E1}\\
&&\nonumber\\
E_2(r,w)&=&(\frac{1}{gT_0}-\alpha\frac{C\Omega}{N(1-\Omega)})-A_2=0
\label{E2}
\end{eqnarray}
which extend and interpolate the results of \cite{Tre+91} to
finite values of $C/N$.

The first equation above appears as a closed curve in the $(w,r)$,
plane, which shrinks in size when one increases $\alpha$ and then
disappears; whereas the second equation is an almost straight
curve, which for a certain range of $g$ intersects twice with the
closed curve above. Since for a given value of $\alpha$ such that
the first equation is satisfied, there always exists a value for
$g$ that satisfies the second equation, the storage capacity is
the value of $\alpha$ for which the closed curve shrinks to a
point. We treat $g$ as a free parameter, because it can be easily
changed in a network by mechanisms like multiplicative inhibition,
if required in order to approach the optimal storage load.

\begin{figure}[h]
\centerline{\hbox{\epsfig{figure=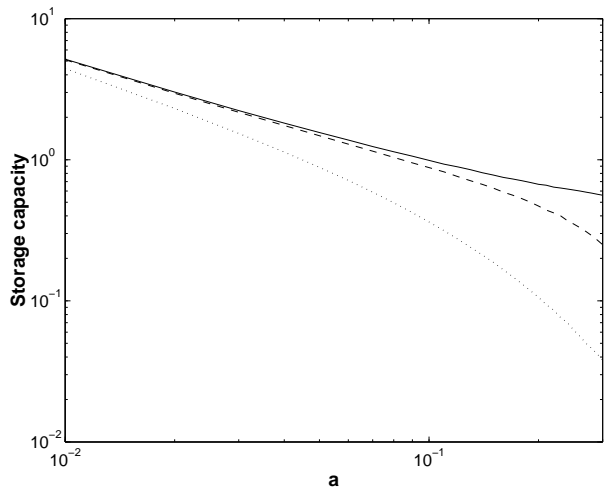,width=8cm,angle=0}}}
\caption{Storage Capacity vs. $a$ for $C/N = 0$ (full curve), $C/N = 0.05$ (dashed line)
 and $C/N=1$ (dotted line). } \label{str}
\end{figure}

In the limit of extreme dilution, \emph{i.e.} $C/N \rightarrow 0$,
$\Omega$ does not contribute to the equation for the storage
capacity. The result of calculating the storage capacity as a
function of the sparseness of the coding is shown in Fig.\ref{str}
(the full curve). For other values of $C/N$ the contribution from
$\Omega$ should be taken into account, which for small $C/N \neq 0$
results in deviations from the storage capacity of a highly diluted
network. An example is illustrated in in Fig.\ref{str} for
$C/N=0.05$. It is clear that, at least for small $a$, a network with
5\% connectivity can be considered as highly diluted, in the sense
that for sparse patterns of activity, the effect of loops -- which
produces the difference between $A_2$ and $A_1$ -- becomes
unimportant.

\section{The network with geometrical connectivity}
\label{geom_network}
In this section we study the fixed points equations of the
network, when the probability of the existence of a connection
between two units depends on their distance, as opposed to the
previous case. An interesting example, in one dimension, is a
network with a Gaussian connectivity probability distribution:
\begin{equation}
Pr\{c_{ij}=1\}=\frac{C}{\sqrt{2 \pi\sigma^2}}e^{\frac{-(i-j)^2}{2\sigma^2}}+ {\rm Baseline}.
\end{equation}
The baseline is a correction that has to be considered for
$\sigma\varpropto N$, to ensure that when the ring cannot be taken
to be infinite, we still have that the sum $\sum_j Pr \{c_{ij}=1\}=C$.

In this {\em geometrical} case there can be spatially non-uniform
solutions to the steady state equations. We have to analyze then
two related issues, both emerging with decreasing $\sigma$, as the
network approaches a more local connectivity: the appearance of
non-uniform solutions to the equations, and whether the storage
capacity of the network decreases.

\subsection{Appearance of spatially non-uniform activity}
\begin{figure}[h]
\centerline{\hbox{\epsfig{figure=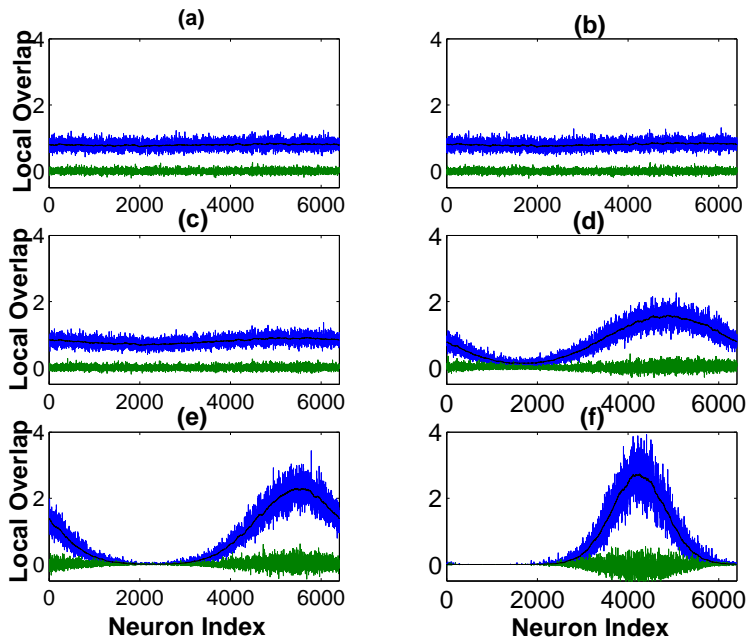,width=10cm,angle=0}}}
\caption{The dependence of the local overlap on $\sigma$. Results
are from simulating a network with $N=6400$, $C=320$, $p=32$,
$g=0.7$, $a=0.2$ and $\sigma=1900$ (a), $\sigma=1700$ (b),
$\sigma=1500$ (c), $\sigma=1000$ (d), $\sigma=700$ (e) and
$\sigma=500$ (f). In each panel, the upper fluctuating curve is
the local overlap with the retrieved pattern and the lower one the
local overlap with one of the non retrieved patterns. The black
line inside the local overlap with the retrieved pattern is a
smooth version of the local overlap, calculated by averaging over
100 nearby units for each point. A smooth change of the local
overlap from a uniform shape to the first Fourier mode is clear.}
\label{sig-dep}
\end{figure}

When $\sigma$ is large one may expect the solutions of the
equations that we have previously found to be again independent of
space. It can be seen from the simulations that this is actually
the case. Indeed when one measures the local overlap with
different patterns in a steady state, one can observe that for
values of $\sigma$ larger than a critical value $\sigma_c$ the
solutions do not show spatial dependence; but, as soon as $\sigma$
becomes smaller than $\sigma_c$, the local overlap begins to
display some spatial dependence, which increases by further
decreasing $\sigma$. This is illustrated in Fig.\ref{sig-dep}. In
this subsection we aim to study this phenomenon analytically.

The simulations indicate that close to the transition the spatial
dependence of the overlap takes the form of a cosine, whereas for
lower $\sigma $ values it approaches a gaussian. One can easily
check analytically, however, that considering a gaussian ansatz
for $m_i$, and another gaussian for, say, $\rho_i$, does not solve
the mean-field equations, Eqs.\ref{mean_field}, which do not in
fact appear to admit any simple curve as a solution. This led us
to develop approximate treatments that circumvent the lack of a
closed-form spatially-dependent solution.

From what we see in the simulations we assume that the transition to
the spatially non-uniform solution is smooth (second order). We
further assume that $C/\sigma_c$ is small so that to a first
approximation, in order to determine the critical point,
we can neglect the effect of loops, \emph{i.e.} of
$\psi$ and $\Gamma$. This assumption may well be inappropriate (for
small $g$, for instance, as we shall see later) but we hypothesize
that its effect will not distort a qualitative picture of the
phenomenon too much. This can be verified by simulations. Using this
ansatz, we now write the solutions of the fixed point equations as
follows:
\begin{eqnarray}
m_i&=&m^0+\delta m_i,\hspace*{0.7cm} \mid\delta m_i \mid\ll m^0\nonumber\\
\rho_i&=&\rho^0+\delta \rho_i,\hspace*{1cm} \mid\delta
\rho_i\mid\ll \rho^0\nonumber\\
T&=& T^{0}+\delta T\hspace*{1cm} \mid\delta T\mid\ll T^{0}
\nonumber
\end{eqnarray}
where $m^0$ and $\rho^0$ are the uniform parts of the solutions,
which we take to be the solutions of the fixed point equations for
$\sigma = \infty$; and $\delta m_i$ and $\delta \rho_i$ are the
small deviations from uniformity. In the same way $T^{0}$ is the
value of the threshold which sets the mean activity $x=a$ for the
uniform solution and $T^{0}+\delta T$ is the (uniform) threshold
necessary to keep $x=a$ in the presence of non-uniform terms
$\delta m_i$ and  $\delta \rho_i$. It is worth noting that a more
accurate approach would be to use the values of $m^0$ and $\rho^0$
calculated for $\sigma$ just above the transition value
$\sigma_c$. These values would be different from those at $\sigma
= \infty$, as the effects of loops may become important close to
the transition to non-uniform solutions; but as stated before we
provisionally neglect this inaccuracy. Using these assumptions and
expanding the mean-field equation around the uniform solutions one
obtains equations for the fluctuations. These equations in the
continuum limit \footnote{The continuum limit can be approached by
averaging both sides of the above equations over a length scale
$\Lambda$ which is large enough to effectively sample the
distributions of both $\{\eta_j\}$ and $\{c_{ij}\}$.} are of the
following form: {\footnotesize
\begin{eqnarray}
\delta m\left(r\right)&=&\int dr' \{\left(\frac{a_{11}}{C}c\left(r,r'\right)-\frac{b_{11}}{N}\right) \delta m\left(r'\right)+ \left(\frac{a_{12}}{C} c\left(r,r'\right)-\frac{b_{12}}{N}\right)\delta \rho\left(r'\right)\}\\
\delta \rho \left(r\right)&=&\int dr' \{\left(\frac{a_{21}}{C}
c\left(r,r'\right)-\frac{b_{21}}{N}\right) \delta
m\left(r'\right)+\left(\frac{a_{22}}{C}c\left(r,r'\right)-\frac{b_{22}}{N}\right)
\delta \rho\left(r'\right)\}
\end{eqnarray}}
where the coefficients are defined in Appendix III.

Next we take the Fourier transform of the two sides of the above
equations, to get:
{\footnotesize
\[\left( \begin{array}{ccc}
1-a_{11}e^{-\frac{k^2\sigma^2}{2}}+b_{21}\frac{\delta(k)}{N} & a_{12}e^{-\frac{k^2\sigma^2}{2}}-b_{12}\frac{\delta(k)}{N}\\
a_{21}e^{-\frac{k^2\sigma^2}{2}}-b_{21}\frac{\delta(k)}{N} & 1-a_{22}e^{-\frac{k^2\sigma^2}{2}}+b_{22}\frac{\delta(k)}{N}
\end{array}\right)
\left(\begin{array}{ccc}
\delta \tilde{m}\left(k\right)\\
\delta \tilde{\rho} \left(k\right)
\end{array}\right)=0.\] }

The above equations for $\delta m$ and $\delta \rho$ have a
non-trivial solution if and only if the determinant of the matrix
of coefficients becomes zero. For $k=0$ the matrix (which includes
the $b$ terms) is the same as that which determines the stability
of the uniform solution in the network without geometry. On the
other hand, when $k\neq 0$ the matrix does not include the $b$
terms, and the vanishing of its determinant yields the critical
value of $\sigma$, as it signals the instability of the uniform
solution towards the appearance of a non-uniform Fourier mode. For
large $\sigma$, the determinant of the coefficient matrix
approaches 1, and it decreases with decreasing $\sigma$. For those
values of the other parameters (besides $\sigma$) for which the
determinant of the matrix is negative at $\sigma=0$, there would
be a critical value for $\sigma$ at which the determinant becomes
zero, and therefore a transition occurs. It is clear that the
first Fourier mode \emph{i.e.} $k=\frac{2\pi}{N}$ is the one that
appears first. This is what one actually observes in simulations.

\begin{figure}[h]
\centerline{\hbox{\epsfig{figure=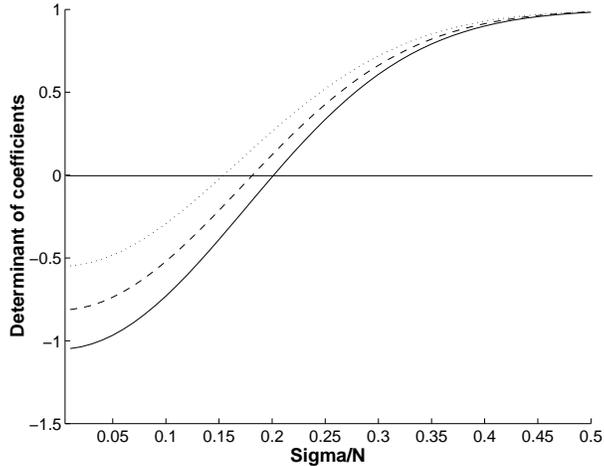,width=8cm,angle=0}}}
\caption{The determinant of the coefficients versus the $\sigma/N$
for $g=0.7$ (full curve), $g=0.6$ (dashed curve) and $g=0.5$
(dotted curve). The other parameters are $a=0.2$ and $p/C=0.1$.}
\label{det}
\end{figure}
In Fig.\ref{det} (full line) we have plotted the determinant of
the coefficient matrix as a function of $\sigma/N$ for the first
Fourier mode. This is done for three values of $g=0.7,0.6$ and
$0.5$. We deduce from this graph that the value of $\sigma_c$
increases by increasing $g$. Note however that the first Fourier
mode, a cosine, is not strictly speaking a solution of
Eqs.\ref{mean_field} for any $\sigma$ value below $\sigma_c$. The
discrepancy becomes more evident, see Fig.\ref{sig-dep}, as
$\sigma$ decreases, and the solution eventually becomes localized,
\emph{i.e.} it takes non-zero values for the order parameters only
on a limited fraction of the ring. In that regime solutions look
like gaussian curves, but they are not exact gaussians.

It is important to realize that the $b$ terms in the above
equations come from the effective threshold which results from our
uniform inhibitory term, and fixes the mean activity of the
network at $x=a$. This is important because if there were no $b$
terms, the condition for the instability of the uniform solution
for the non-geometric network and the condition for the appearance
of the non-uniform solutions at finite $\sigma$ would have been
the same. Therefore without an activity-dependent threshold,
stable retrieval in the non-geometric network would have implied
no spatially non-uniform solution in its geometric counterpart.

\subsection{Storage capacity}
The storage capacity of the geometric network differs from that of
the non-geometric one for two main reasons. The first one is that by
changing the geometry of the connectivity one changes the
distribution of the connectivity loops that contribute to noise
reverberation in the network. Effectively, lowering $\sigma $
increases the \emph{clustering} of the nodes in the network
\cite{Wat+98} and the number of its loops, leading to a decrease in
storage capacity for the same reason that the capacity decreases (if
expressed as $\alpha\equiv p/C$) when a diluted non-geometric
network reaches a denser connectivity.

The second one is the non-uniformity of the spatial distribution
of the signal and noise to the units i.e. the spatial dependence
of $m(r)$ and $\rho(r)$. Effectively, the connections originating
from the less active units at the flanks of the spatial profile
are used less, or even unused if the solution is localized and
those units remain inactive, and the network becomes roughly
equivalent to one with a lower $C$ value.

These two effects are correlated with each other, but
one can get an estimate of how they affect the storage capacity by
first considering them separately. In other words, one can consider a
network with structured connectivity and calculate its storage
capacity if the profile of activity has no spatial dependence.
Although we know from the previous sections that this kind of
uniform solution would not be the stable state of the network, we
can calculate this way the effect of the change in the
distribution of the loops on the storage capacity. On the other
hand, one can consider a network without geometric connectivity,
but with a spatially non-uniform activity, although again this
would not be the stable solution. We will follow this approach in the
coming subsections.

\subsubsection{The effect of the loops}
We start by considering a uniform solution to the mean-field equations Eqs.\ref{mean_field},
while considering a gaussian connectivity. The uniform solution is
of course not stable nor relevant when $\sigma < \sigma_c$: as discussed previously (see Fig.\ref{sig-dep} and section 4.1)
the actual form of the activity is quite close to a cosine immediately
below the $\sigma_c$; and then slowly transforms to a localized quasi-gaussian activity (Fig\ref{sig-dep}.f) by decreasing $
\sigma$. Although we know that the uniform solution is not stable below $\sigma_c$,
we can still calculate a reference storage capacity by inserting it into the mean-field equations. Then we can compare the results
with the simulations and assess their correspondence. In the next section we shall see that
this procedure, even though very crude, gives us an estimate of the true storage capacity which
is comparable with the simulations.

By considering the uniform solution to the equations, we have:
\begin{equation}
\psi(r-r')=\frac{C}{\sqrt{2\pi
\sigma^{2}}} \sum_{n=1}^{\infty}e^{-\frac{(r-r')}{2(n+1)\sigma^{2}}}
\frac{1}{\sqrt{n+1}}\Omega^{n}
\end{equation}
 and hence:
\begin{equation}
\int dr''
\{2c(r-r'')\psi(r-r'')+\psi(r-r'')^{2}\}=\frac{C}{\sqrt{2\pi
\sigma^{2}}}\sum_{n=1}^{\infty} \frac{n+1}{\sqrt{n+2}}\Omega^{n}
\end{equation}

Using the above, we can write the equation for the storage
capacity as:
\begin{equation}
A_2-\left(1+\frac{C}{\sqrt{2\pi \sigma^{2}}}\sum_{n=1}^{\infty}
\frac{n+1}{\sqrt{n+2}}\Omega^{n}\right)\alpha A_3=0
\end{equation}
Using this equation and evaluating the series numerically, we have
calculated the storage capacity for various values of $\sigma$ in
a network of $C=320$ and $a=0.2$, as shown in Fig.\ref{str_geom}.
This graph indicates that the storage capacity decreases with
$\sigma$ but not by much. The assumption of considering the
uniform solution implies that such a decrease is due
solely to the increased relevance, as $\sigma$ decreases, of
closed loops, and therefore of increased noise reverberation.
Although this analysis does not take into account additional
effects due to the emergence of non-uniform solutions, we shall
see in the next section that the capacity decrease in the graph is
quite comparable with the results of the simulations, also for
values of $\sigma$ which lead to non-uniform steady states.

\subsubsection{The effect of the non-uniform solution}
We can have an idea of how the form of the solution affects the
storage capacity simply by considering an ansatz on the form of
$\rho(r)$ and $m(r)$ which depends on a finite set of parameters
$\{\lambda_1,\lambda_2,\cdots,\lambda_k \}$. We also assume, to start with, that
the effects of loops are negligible, so that we can set
$\psi(r,r')=0$ and $\Gamma(r)=0$. Loops are of course not negligible {\em e.g.} when $\sigma$
is small, but we want now to isolate the effect of the non-uniformity from that of loops,
which has been estimated in the previous part using the uniform solution.
In other words by this procedure we actually consider a network without loops and calculate
the storage capacity corresponding to a parametrized non-uniform activity profile,
although we know it is unstable in a structure-less network.

With this ansatz and integrating over $r$ on both the left and right hand sides of the
equations describing $\rho(r)$ and $m(r)$, i.e.
Eqs.\ref{mean_field}, we can write after some manipulation:
\begin{eqnarray}
\left( \frac{\int dr I_2(r)}{T_0\int dr m(r)}\right)^2&-&\alpha
\frac{\int dr I_3(r)}{\int dr [\rho(r)]^2}\label{str_geom_eq}=0\\
x&=&\frac{\int dr m(r)}{\int dr I_2(r)}\int dr I_4(r)
\label{activ}
\end{eqnarray}

\begin{eqnarray*}
I_2(r)&=& \langle(\frac{\eta(r)}{a}-1) \int^+ Dz
((\frac{\eta(r)}{a}-1)m(r)+b(x)-T_{thr}-\rho(r)z)\rangle\\
I_3(r)&=&\langle \int^+ Dz
((\frac{\eta(r)}{a}-1)m(r)+b(x)-T_{thr}-\rho(r)z)^2\rangle\\
I_4(r)&=&\langle\int^{+} Dz
((\frac{\eta(r)}{a}-1)m(r)+b(x)-T_{thr}-\rho(r)z)\rangle
\end{eqnarray*}
In practice, for any given form of the functions $m(r)$ and
$\rho(r)$, one can solve Eq.\ref{activ} to get
$b(x)-T_{thr}$ by setting its r.h.s to $x=a$. Then one can use this to
evaluate the integrals appearing in Eq.\ref{str_geom_eq} and find
the highest value of $\alpha$ for which a solution for this equation exists
in the $\{\lambda_1,\lambda_2,\cdots,\lambda_k \}$ space.

As an example, let us consider a gaussian form, which simulations indicate
is a reasonable approximation, even if not an exact solution, in the
localized, or low $\sigma $, regime. In other words, let us assume that
$m(r)= m_0 exp(-r^2/2l^2)$ and $\rho(r)=\rho_0
exp(-r^2/2l^2)$.
By fixing $l$ and following the procedure
described above one finds that for a given value of $\alpha$,
Eq.\ref{str_geom_eq} appears as a closed curve in the
$(m_0,\rho_0)$ plane, that shrinks in size with increasing $\alpha$, analogously to Eq.\ref{E1}.
The value of $\alpha$ for which this closed curve disappears
defines the storage capacity at constant width $\alpha(l)$.
Of course the gaussian ansatz is an approximation which needs to be considered carefully.
An example of gaussian-like activity is Fig.\ref{sig-dep}.f.
Unfortunately the profile of activity below $\sigma_c$ does not take an
analytical form. Still, a gaussian
fit seems to be a good approximation, even though it is not the solution
(note that in Fig.\ref{sig-dep}.f the activity goes to zero outside a finite radius, hence it cannot be a gaussain).
The accuracy of the gaussian ansatz can be checked by comparing the resulting storage capacity with that
of the simulations. As we shall discuss in the next section, our procedure leads
to a reasonable agreement with the simulations.

\begin{figure}[h]
\centerline{\hbox{\epsfig{figure=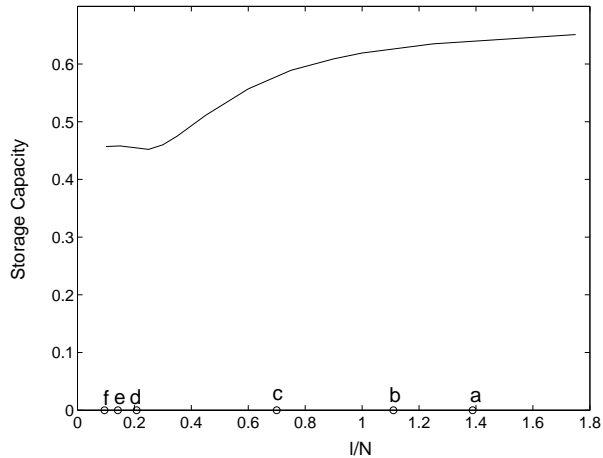,width=8cm,angle=0}}}
\caption{The dependence of the storage capacity, for $a=0.2$, on
$l/N$, where $l$ is the width of a gaussian solution. Note the correspondence
with the $\sigma$ values used in Fig.\ref{sig-dep}: see text. }
\label{alpha_l}
\end{figure}

Fig.\ref{alpha_l} shows the result of following the above procedure
for calculating $\alpha(l)$ for a network of $N$ units and $a=0.2$.
The decrease in the storage capacity for more localized solutions can be seen.
This decrease is solely due to the non-uniformity of the solution, and it has no
contribution from the geometry of the connectivity, since the
effect of the connectivity decouples when one integrates over
space in the mean-field equations to get Eq.\ref{str_geom_eq}.
In other words, there is no dependence on the connectivity probability distribution $c(r,r')$ in this equation.
In Fig.\ref{alpha_l} we have indicated, on the $l/N$ axis, the
widths of the profiles seen in the simulations with the 6 values of
$\sigma $ used in Fig.\ref{sig-dep}. We have run extensive simulations with those 6 values of
$\sigma$ but, unlike those in Fig.\ref{sig-dep}, with $\alpha $ close to $\alpha_c$ and $g$
to the optimal gain value, as estimated from the simulations themselves. To calculate the
values of $l$ corresponding to each $\sigma$ and close to capacity we have used equation
Eq.~\ref{q-sim}: we first calculate the average value of $q$ across simulations with
each $\sigma$, by using its definition Eq.\ref{q}, and then find the value $l$ which
solves Eq.\ref{q-sim}. Please note that the gaussian fit is a reaonable approximation,
only in the localized regime {\em e.g.} Fig.\ref{sig-dep}e and f,
but not really for apparently flat solutions like those in Fig.\ref{sig-dep}a or even
cosine-like ones like those in Fig.\ref{sig-dep}c. Moreover, finite size effects smooth
the otherwise sharp transition at $\sigma_c$. Finally, Fig.\ref{sig-dep} was produced
by running simulations at low $\alpha$ and fixed $g$, whereas we now set these parameters
at the storage capacity limit. All these effects cumulate to make the estimated $l$ values
smaller than what one would have predicted from visually inspecting Fig.\ref{sig-dep}a-c
(the discrepancy is milder in the localized regime). For example, looking at point $a$
in Fig.\ref{alpha_l} one sees $l=1.4 N$ for $\sigma =1900$, while from the flat-looking
solution of Fig.\ref{sig-dep} one might have expected $l\to\infty$!

One can see that such widths change significantly, and correspondingly
there is a significant estimated capacity decrease, for the upper 3 $\sigma$ values,
which are relatively clustered around $\sigma_c$. For the lower 3
$\sigma$ values, even though they cover a much larger range on a
log scale, the resulting profile widths change less ($l$ is
roughly proportional to $\sigma$), as the solutions have become
effectively localized. Correspondingly, the storage capacity does
not decrease further due the profiles of the solutions, although
it continues to decrease due to more clustered connectivity. The
second of the two mechanisms which decrease the storage capacity
thus reaches a maximum effect as soon as the two flanks of the
activity profile go to zero, and the steady state of the network
is a genuine 'bump'.

\section{Simulations}
In this section we present the results of simulations that
investigate the relation between the storage capacity and the
width of the connectivity, as well as the emergence of non-uniform
asymptotic states. Such states for very local connectivity (very
low $\sigma $) eventually become {\em localized}, in the sense
that activity is zero outside a limited fraction of the ring.

To measure the degree of uniformity of the steady states reached
in each simulation, we define the quantity:
\begin{equation}
q=\frac{12\int dr
\left(r-r_{max}\right)^{2}m\left(r\right)}{N^2\int dr
m\left(r\right)}
\label{q}
\end{equation}
where $r_{max}$ is where the local overlap $m\left(r\right)$ has
its maximum. We use a smoothed version of the local overlap to
regularize the parameter $0 < q \le 1$, which takes its maximum
$q=1$ for a uniform solution and is inversely related to the
degree of {\em bumpiness}, or of locality, of a spatially
non-uniform overlap distribution \cite{Ani+04}.

\begin{figure}[h]
\centerline{\hbox{\epsfig{figure=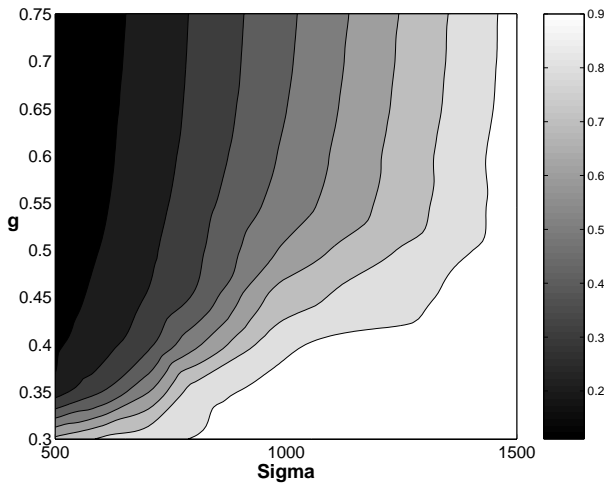,width=8cm,angle=0}}}
\caption{The change in the uniformity $q$ as a function of $g$ and
$\sigma$ from simulating a network of $N=6400$, $C=320$. The
sparsity of the patterns is $a=0.2$. The value of $p$ is chosen in
a way that the network on average is able to retrieve $50 \%$ of
the patterns for the given values of $g$ and $\sigma$.}
\label{g_sigma}
\end{figure}

Here we report the results of simulating a network composed of
$N=6400$ units with $C=320$ and $a=0.2$. For each value of
$\sigma$, $g$ and $p$ we give the network a full cue,
corresponding to one of the stored patterns, and after $50$
synchronous updates we measure the final overlap with the
presented pattern. If the final overlap is larger than $0.4$ we
take it to be a successful retrieval. We repeat this for $4$
different seeds for the random number generator and $5$ different
patterns and then index the performance of the network with the
percent of patterns retrieved (according to the above criteria).
We take the value of $p$ at which the performance reaches $50\%$ as a
measure of the storage capacity for those values of $g$ and
$\sigma$. By repeating for different values of $g$ we thus assess
the {\em optimal} storage capacity for the network, optimized
across values of the gain.

In Fig.\ref{g_sigma} one can observe the way $q$ changes as a
function of $g$ and $\sigma$. As we found in section
\ref{geom_network}, for values of $\sigma$ below the transition
the activity is not uniform in space, i.e. it has a 'bumpy'
profile. Increasing $g$ favours the localization of the solution,
which for high $g$ and low $\sigma$ becomes a genuine 'bump' of
activity, as seen in Fig.\ref{sig-dep}. There appears to be no
sharp transition, but rather a very smooth cross-over, from the
'quasi-cosine' regime near $\sigma_c$ to the localized
'quasi-gaussian' regime for low $\sigma$. The cross-over is in
fact regulated by the gain $g$.

\begin{figure}[h]
\centerline{\hbox{\epsfig{figure=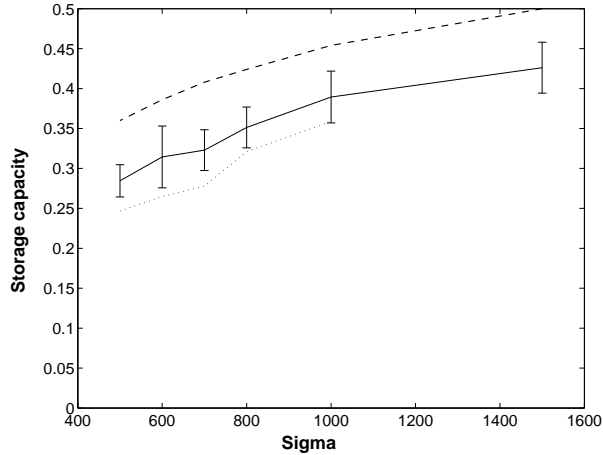,width=8cm,angle=0}}}
\caption{In this graph, the full curve shows the storage capacity
versus $\sigma$ for $a=0.2$, obtained numerically simulating a network with
$N=6400$, $C=320$. The dashed curve is the storage capacity
estimated analytically using the uniform approximation, as discussed
in section 4.2.1. The dotted line represents the storage capacity
estimated taking into account, as uncorrelated, both the effect of more
loops and that of the spatial dependence of the solution, as explained in section 4.2.2.}
\label{str_geom}
\end{figure}

The second point that we have studied through the simulations is
the dependence of the storage capacity on the width of the
connectivity $\sigma$. The storage capacity calculated from the
simulations is the full curve in Fig.\ref{str_geom}. This curve
lies below the storage capacity calculated analytically using the
uniform solution. As previously mentioned there are two effects
which contribute to the decrease of the storage capacity. First, the
increase in the number of loops as a result of the decrease in the width of the
connectivity, which has the effect discussed in section 4.2.1 and shown with
the dashed curve in Fig.\ref{str_geom}. Second, the non-uniform profile of
the solution, discussed in section 4.2.2. These two effects are not
uncorrelated but one can consider them combined as if they were uncorrelated, in the following way.
For each value of $\sigma$ one first calculates the storage capacity by considering only the effects
of the loops, i.e. the dashed curve. Then one estimates the most appropriate width for a
model localized solution, i.e. the best gaussian fit,
by solving the following equation for $l$, using the value $q$ obtained from
the simulations:
\begin{equation}
q=\frac{12\int_{-0.5}^{0.5} dr r^{2}
exp(-\frac{r^2N^2}{2l^2})}{\int_{-0.5}^{0.5} dr
exp(-\frac{r^2N^2}{2l^2})}
\label{q-sim}
\end{equation}
Then an estimate of the storage capacity, given independent effects of loops and localization, would simply be the
multiplication of the storage capacity calculated for the uniform
solution by the factor $\alpha(l)/\alpha(\infty)$. This is the dotted
curve in Fig.\ref{str_geom}. It yields a lower estimate of
the storage capacity as compared to that of the simulations, but
yet closer to it than the uniform solution approximation. The assumption of
uncorrelated effects thus overestimates the capacity decrease with lower $\sigma$ values.

Note that the capacity decrease could in principle be overestimated also
as an effect of our procedure of exploring only spatially-dependent solutions
of a given shape (in particular, gaussian); this effect is however likely negligible,
at least in the localized regime where the profiles seen in the simulations are very close to gaussian.

\section{Discussion}
In this paper we have studied the retrieval properties of an
autoassociative network with geometric synaptic connectivity. To
approach the analysis of a spatially structured network we had to
find alternatives to the standard thermodynamic formalism often
applied to systems with quenched disorder, based on the calculation
of a free-energy with the replica trick, and on its evaluation at a
saddle point. Even though in terms of their behaviour and even of
their fixed point equations asymmetric and symmetric networks do not
differ by much at least in networks with threshold-linear units for
which the spin glass phase is irrelevant \cite{Tre91b}, the lack
of a Hamiltonian requires an alternative approach to obtain the
mean-field equations. Fukai and Shiino developed years ago a
'self-consistent signal-to-noise analysis' to treat cases in which a
Hamiltonian could not be defined, in particular asymmetric networks
\cite{Shi+92} and networks with arbitrary analogue transfer
functions \cite{Shi+93}. The asymmetric connectivity is in some
sense a technical problem, which might not reveal any 'new' physics
in the threshold-linear networks. In our case, we had to face a
second and more substantive problem, as a direct result of the {\em
geometry} in the connectivity. This is the fact that in order to
describe the behavior of a geometric network, one needs to introduce
order parameters that are not scalars but scalar {\em fields}. This
makes the derivation of the mean-field equations and the analysis of
their solutions much more challenging: the equations are now
integral equations in these field order parameters (see
Eq.\ref{mean_field}). We adapted the 'self-consistent
signal-to-noise analysis' turning it into a {\em local}
signal-to-noise approach, to find self-consistent equations for
order parameters with the following physical meaning:
\begin{itemize}
\item $m(r)$: the {\em local overlap}, i.e. the product of the
current activity and the stored patterns, summed over all units
presynaptic to the unit at $r$.
\item $\rho(r)$: the {\em local noise}, i.e. the mean square amplitude
of the non-condensed overlaps \cite{Ami89}, as seen at $r$.
\item $\psi(r,r')$: the effect of the direct and indirect connections
linking two units at positions $r$ and $r'$ on the reverberation of the noise.
\item $\Gamma(r)$: a measure, proportional to the diagonal elements of
$\psi(r,r')$, of the effect of the activity of each unit on the
noise component of its own input.
\end{itemize}

Although the stored patterns do not include any spatial structure
or correlation, the retrieval states of the geometric connectivity
network, as simulations easily demonstrate, may have non-uniform
activity profiles when the connections are short range enough. The
activity profile of the retrieval state can even become localized
in space, as a 'bump' of activity, in the appropriate parameter
regime. It is worth emphasizing that a non-uniform retrieval state
does not correspond in full to the stored pattern that is being
retrieved, since the stored pattern does not have any spatial
preference; the retrieval state has a large overlap with the
pattern, but circumscribed to the bump. As shown in
Fig.\ref{g_sigma}, the bumpiness of such a non-uniform retrieval
state depends not only on how short range is the connectivity, but
also on the gain of the input-output transfer function. Increasing
the gain favours localization in the retrieval state, of course as
long as the network remains in the regime where retrieval occurs.

Given that the storage capacity of the geometric network decreases
with respect to that of a uniform network due to both an increase
in the loops and the non-uniformity of the solution, we described
a procedure to estimate these effects under the assumption that
they are independent. Comparing with simulations, we found that
the procedure yields a reasonable estimate of the storage
capacity, although lower than the true value, due probably to the
simplifying independence assumption. The bottom-line conclusion is that as the
connectivity of the autoassociative network becomes shorter range,
its storage capacity, expressed as $p/C$, decreases indeed, but
not by a large factor.

The model that we have studied here is defined on a ring, i.e. it
is a one dimensional model. Although we do not expect significant
differences between the results reported here and those of a more
interesting model in two dimensions, it remains to be checked, in
future work, to what extent the results can be generalized to a 2D
model.

Further, we have considered a gaussian connectivity of varying
width. An alternative model, introduced by Watts and Strogatz
\cite{Wat+98}, is that of a network with a fraction of connections
strictly short range, and the complementary fraction distributed
at random across the entire network. This could be perhaps an
appropriate model for cortical connectivity, with its short- and
long-range connections \cite{Bra+91}, and it has already been
considered, albeit in a geometry-free formulation, as a model of
semantic memory \cite{O'K+92,Ful+98}. It seems important to be
able to extend our approach to a model of the Watts-Strogatz type,
with 1D or 2D geometry. It is possible that the storage capacity
would decrease more substantially in such a model, than it does in
the one considered here - a naive expectation is that $p$ should
scale, essentially, with the number of short range connections, those
that comprise the so-called {\em regular} network.

Indeed, the Watts-Strogatz autoassociative model has been studied
by \cite{Ani+04}, even though with simulations only and a more
realistic neuronal integrate-and-fire dynamics. One result of that
study is the near incompatibility between localization and
retrieval, which appear to occur in almost mutually exclusive
ranges of the relevant parameter, the fraction of short-range
connections. The fact that retrieval did not seem to succeed when
asymptotic firing states are localized may have been due to a
strongly decreased capacity, or to instabilities associated with
the integrate-and-fire dynamics, or to other aspects of that
model, such as the mechanism implementing inhibitory control of
the activity of excitatory units. Whatever the case, this
emphasizes the need for studies of associative retrieval and
localization in realistic neural network models. Important
features that need consideration are the dynamics of individual
units and the model adopted for inhibitory effects.

Once thus extended, this approach has the potential to yield a
quantitative assessment of the storage capacity of cortical
modules in the mammalian brain. Such modules obviously differ from
our simplified mathematical models in several respects, but an
educated guess predicts that the crucial factor that determines
their storage capacity is the number and geometrical distribution
of the Hebbian-modifiable connections that the average unit
receives.

It becomes possible at that stage to consider in a realistic setting
a phenomenon which had been considered earlier in rather abstract models:
the coexistence of multiple retrieval states. So-called 'spurious' solutions
have been shown not to be stable, essentially, in threshold-linear networks
without geometry \cite{Rou+03}, but in a simple ring model \cite{Abr+04}
different stored patterns can be retrieved in different portions of the ring.
Thus the underlying geometrical manifold can turn spurious mixtures of
patterns into interesting combinations of localized patterns, and this
raises the issue of their competitive interactions. Recent neurophysiological
work \cite{Rol+04} demonstrates that the receptive fields of visually evoked
activity patterns are effectively restricted 2- to 3-fold, in the macaque
temporal cortex, when several objects are present together in the visual scene.
This finding may have a correlate in an associative memory network with
geometry, where long-range inhibition may restrict the width of activity
profiles retrieved on different portions of the manifold.


\renewcommand{\theequation}{I-\arabic{equation}}
\setcounter{equation}{0}
\section*{Appendix.I}
Using Eq.\ref{mimu}, the solution for $m^{\mu}_i$ can be written
as:
\begin{equation}
m_i^{\mu}=\frac{1}{C}R^{\mu}_{ii}(\eta_i^{\mu}/a-1)G^{\mu}[i]+\frac{1}{C}\sum_{j\neq
i}R^{\mu}_{ij} (\eta_j^{\mu}/a-1)G^{\mu}[j]
\end{equation}
where $R_{ij}^{\mu}$ is expanded in a power series as:
\begin{equation}
R_{ij}^{\mu}=c_{ij}+\sum_l K_{il}^{\mu}c_{lj}+\sum_{lt}
K_{il}^{\mu}K_{lt}^{\mu}c_{tj}+\ldots
\end{equation}
and we have used the notation $G^{\mu}[i]=G[(\eta_{i}^{1}/a-1)
m_{i}^{1}+\rho_{i}^{\mu}z+b(x)-T_{thr}]$.

Now that we have expressed the local overlap with a
non-retrieved pattern as a function of $m^{1}_i$, we can proceed
to evaluate $\rho$ and $\gamma$:
\begin{eqnarray}
\sum_{\nu \neq 1}(\eta_i^{\nu}-1)m_i^{\nu}&=&\frac{1}{C}\sum_{\nu \neq 1} R_{ii}^{\nu}(\eta_i^{\nu}/a-1)^2 G^{\nu}[i]\label{eq15}\\
&+&\frac{1}{C}\sum_{j\neq i,\nu \neq
1}R_{ij}^{\nu}(\eta_i^{\nu}/a-1)(\eta_j^{\nu}/a-1)G^{\nu}[j].\nonumber
\end{eqnarray}

For the first sum in the r.h.s of Eq.\ref{eq15} above, using the
independence of different patterns and assuming that
$\rho_i^{\mu}\simeq \rho_i$ one can write:
\begin{eqnarray}
\frac{1}{C}\sum_{\nu \neq 1}R_{ii}^{\nu}(\eta_i^{\nu}/a-1)^2 G^{\nu}[i]&=&\alpha \langle R_{ii}^{\nu}(\eta_{i}^{\nu}/a-1)^2 G^{\nu}[i]\rangle \\
&\simeq&\alpha \langle R_{ii}^{\nu}(\eta_{i}^{\nu}/a-1)^2\rangle
v_i\nonumber
\end{eqnarray}
and as a result we identify:
\begin{equation}
\gamma_i=\alpha <R_{ii}^{\nu}(\eta_{i}^{\nu}/a-1)^2>=\alpha T_0
\langle R_{ii}^{\nu}\rangle,
\end{equation}
where we denote as in \cite{Tre90} $T_0\equiv 1/a-1$; and
therefore:
\begin{equation}
\Gamma_i= \alpha T_0(\langle R_{ii}^{\nu}\rangle-c_{ii})=\alpha
T_0\langle \sum_l K_{il}^{\mu}c_{lj}+\sum_{lt}
K_{il}^{\mu}K_{lt}^{\mu}c_{tj}+\dots \rangle.
\end{equation}

The second term is a bit tricky. For this term, by replacing the
sum with the average we get zero mean, but for the standard
deviation we have:
\begin{equation}
\rho^2=\frac{\alpha}{C}(1/a-1) \sum_{j}
\langle{R_{ij}^{\nu}}^2(\eta_j^{\nu}/a-1)^2 G^{\nu}[j]^2\rangle
\end{equation}
which is, actually, the standard deviation of the noise. We can
then replace the second term, that is the noise term, with a
gaussian random variable with mean zero and standard deviation
$\rho$, and take it into account in our mean-field equations by
averaging the equations over this gaussian measure.
We shall discuss the reliability of this assumption soon.

 In order to make the equations more comprehensive, we define
the order parameter $\psi$ as an average over non-retrieved patterns
of $\psi^{\mu}$, where:
\begin{equation}
\psi_{ij}^{\mu}=R^{\mu}_{ij}-c_{ij}=\sum_l
K_{il}^{\mu}c_{lj}+\sum_{lt} K_{il}^{\mu}K_{lt}^{\mu}c_{tj}+\ldots
\label{loop}
\end{equation}
By using this definition with some algebraic manipulation we get
the mean-field equations.

Considering the noise term as a gaussian random variable comes from
rewriting the left hand side of Eq.\ref{noise} in its original
form and then approximating it as a sum of random independent variables:

\begin{equation}
\rho_i z+\gamma_i v_i=\frac{1}{C}\sum_{\nu \neq
1,j}c_{ij}(\eta_i^{\nu}/a-1)(\eta_j^{\nu}/a-1)v_j
\end{equation}

The assumption of independence of these terms is a bit tricky, since
in general when following the dynamics of the system the activity of each
unit $v_i$ at a given time depends on the local field at the previous
time step, hence on the patterns. However, it becomes an
appropriate approximation when the first pattern is thoroughly retrieved {\em i.e.}
$v_i=\eta_i^1$. In this case the terms that appear
in the noise sum are in fact close to independent random variables, as the patterns are, by construction.
The gaussian noise assumption is thus accurate in the limit in which the solutions of
the mean-field equations are exactly equal to the patterns used in
generating the weight matrix, {\em i.e. } in the case of retrieval
without errors. This is of course never strictly the case, unless the
storage capacity is zero in the thermodynamic limit\cite{Ama+88}.
However, if we assume that the retrieved state in a successful retrieval
is very close to a stored pattern and has a large overlap with it,
then the gaussian approximation is reasonable.
In other words the gaussian noise is appropriate when
retrieval really occurs.

It is important to note that one should be careful in using the
gaussian approximation when dealing with the dynamics of the
network. This approximation may just qualitatively predict the
recall dynamics when retrieval occurs (as well as the stationary
state phase diagram) but it fails to describe the non-retrieval
trajectories. This is simply due to the fact that the noise
distribution can be approximated by a gaussian when retrieval is
successful, as explained above, but it could be non-gaussian in the
intermediate states before getting close to the retrieval attractor.
Starting from an initial state which for instance has a non-zero
overlap with two patterns, the noise does not follow in general a gaussian distribution.
It approaches a gaussian form as the network evolves toward one of the attractors,
corresponding to one of the patterns.
The gaussian approximation may also give the correct dynamical
equations for the fist few time steps,
but then it may give results different from the
exact solutions later on. For a detailed discussion on this issue see
\cite{coo00, Bol+03}.

As we shall see later in the limit of a structure-less network the mean-field equations from the gaussian
approximation are identical to those found previously using the
replica method. This confirms that the equations that result from
the gaussian approximation in the signal-to-noise analysis are in terms of accuracy in the same line
as the equations derived with the replica method. Actually
the close relation between the signal-to-noise analysis, the replica
method and the TAP equation has been investigated in a recent paper
by Shiino and Yamana \cite{Shi04}.


\renewcommand{\theequation}{II-\arabic{equation}}
\setcounter{equation}{0}
\section*{Appendix.II}
Dividing by the gain factor the equations for $\rho$ and $m$ one
gets:
\begin{eqnarray*}
\left(\frac{gT_0}{1-g\Gamma}\right)^{-2}&=&\alpha\left(1+2\psi+\frac{N}{C}\psi^{2}\right)A_3\\
\left(\frac{gT_0}{1-g\Gamma}\right)^{-1}&=&A_2
\end{eqnarray*}
On the other hand using the equation for $\psi$ and the definition
of $\Omega$ in Eq.\ref{Omega_def} we have:
\begin{eqnarray*}
2\psi+\frac{N}{C}\psi^{2}&=&\frac{C}{N}\frac{(2-\Omega)\Omega}{(1-\Omega)^{2}}
\end{eqnarray*}
and we find:
\begin{equation}
\Omega=\langle\int^{+}Dz\rangle/A_2. \nonumber
\end{equation}

Combining the equations above one gets to Eqs.\ref{E1} and
\ref{E2}. The second of those equations determines the optimal
value of $g$, but it does not change the storage capacity.

\renewcommand{\theequation}{III-\arabic{equation}}
\setcounter{equation}{0}
\section*{Appendix.III}
\begin{eqnarray*}
a_{11}&=& g\langle\left(\frac{\eta}{a}-1\right)^2 \phi\left(w+v\eta/a\right)\rangle\\
b_{11}&=& g\langle\left(\frac{\eta}{a}-1\right)\phi\left(w+v\eta/a\right)\rangle^{2}/ \Delta\\
a_{12}&=& g\langle \left(\frac{\eta}{a}-1\right)\beta\left(w+v\eta/a\right)\rangle\\
b_{12}&=& g\langle \left(\frac{\eta}{a}-1\right)\phi\left(w+v\eta/a\right)\rangle \langle \beta\left(w+v\eta/a\right)\rangle / \Delta\\
a_{21}&=& \alpha g^{2}T_0^{2}\langle(\frac{\eta}{a}-1)\left((w+v\eta/a)\phi\left(w+v\eta/a\right)+ \beta\left(w+v\eta/a\right)\right)\rangle\\
b_{12}&=&\alpha g^{2}T_0^{2}\langle\left((w+v\eta/a)\phi\left(w+v\eta/a\right)+ \beta\left(w+v\eta/a\right)\right)\rangle \langle\left(\frac{\eta}{a}-1\right)\phi\left(w+v\eta/a\right)\rangle/ \Delta\\
a_{22}&=& \alpha g^{2}T_0^{2}\langle\phi\left(w+v\eta/a\right)\rangle\\
b_{22}&=&\alpha g^{2}T_0^{2}\langle\left((w+v\eta/a)\phi\left(w+v\eta/a\right)+ \beta\left(w+v\eta/a\right)\right)\rangle \langle\beta\left(w+v\eta/a\right)\rangle/ \Delta\\
\Delta &=& \langle\phi\left(w+v\eta/a\right)\rangle\\
\phi\left(x\right)&=&\frac{1}{\sqrt{2\pi \sigma^{2}}}\int_{-\infty}^{x}dz \exp(-z^2/2)\\
\beta\left(x\right)&=&\frac{1}{\sqrt{2\pi
\sigma^{2}}}\exp(-x^2/2).
\end{eqnarray*}
where we have set $r=m_0/\rho_0$ and
$w=[b\left(x\right)-m_0-T_{thr}]/\rho_0$.

\end{document}